\documentclass[energies,article,accept,moreauthors,pdftex,12pt,a4paper]{mdpi} 

\lastpage{x}
\doinum{10.3390/------}
\pubvolume{xx}
\pubyear{2014}
\history{Received: xx / Accepted: xx / Published: xx}

\usepackage{graphicx,float,amssymb}

\usepackage{color}
\usepackage{soul}                    
\definecolor{red}{rgb}{1,0,0}
\definecolor{green}{rgb}{0,1,0}
\definecolor{blue}{rgb}{0,0,1}

\Title{Fatigue Loads Estimation Through a Simple Stochastic Model}

\Author{Pedro G.~Lind, Iv\'an Herr\'aez, M.~W\"achter, J.~Peinke}

\address{
ForWind - Center for Wind Energy Research, Institute of Physics,\\
Carl von Ossietzky University of Oldenburg, 26111 Oldenburg, Germany}

\corres{Email: pedro.g.lind@forwind.de, Phone: +49(0)4417985051, Fax: +49(0)4417985099}

\abstract{%
We propose a procedure to estimate the fatigue loads on wind turbines, 
based in a recent framework used for reconstructing data 
series of stochastic properties measured at wind turbines.
Through a standard fatigue analysis, we show that it is possible to 
accurately estimate fatigue loads in any wind turbine within one wind park, 
using only the load measurements at one single turbine and the set of
wind speed measurements.
Our framework consists of deriving a stochastic differential equation that 
describes the evolution of the torque at one wind turbine driven by the 
wind speed. 
The stochastic equation is derived directly from the measurements and 
is afterwards used for predicting the fatigue loads at neighboring turbines.
Such a framework could be used to mitigate the financial efforts usually 
necessary for placing measurement devices in all wind turbines within one 
wind farm.
Finally, we also discuss the limitations and possible improvements of the
proposed procedure.}


\begin{document}

\section{From the load at one single turbine to the fatigue of an entire wind farm}

Wind energy has become one of the most promising answers to the
world-wide energetic problem\cite{johnsonWindBook}, profiting from
the recent developments and research activities in engineering,
meteorology and physical sciences. However, important challenges
still remain to be solved, particularly in two fronts. 

First, in what 
concerns the predictability and optimization of power production 
when the wind is itself turbulent and intermittent\cite{davidmuecke},
i.e.~large fluctuations occur with a non-negligible 
probability\cite{patrickprl}.

Second, the costs implied in the construction of wind turbines together 
with the proper devices for controlling and monitoring them are typically
high\cite{windenergyhandbook}. Alternative methods to indirectly estimate
the necessary quantities for control and monitoring could help to mitigate
these costs.
\begin{figure}[t]
\begin{center}
\includegraphics[width=0.85\textwidth]{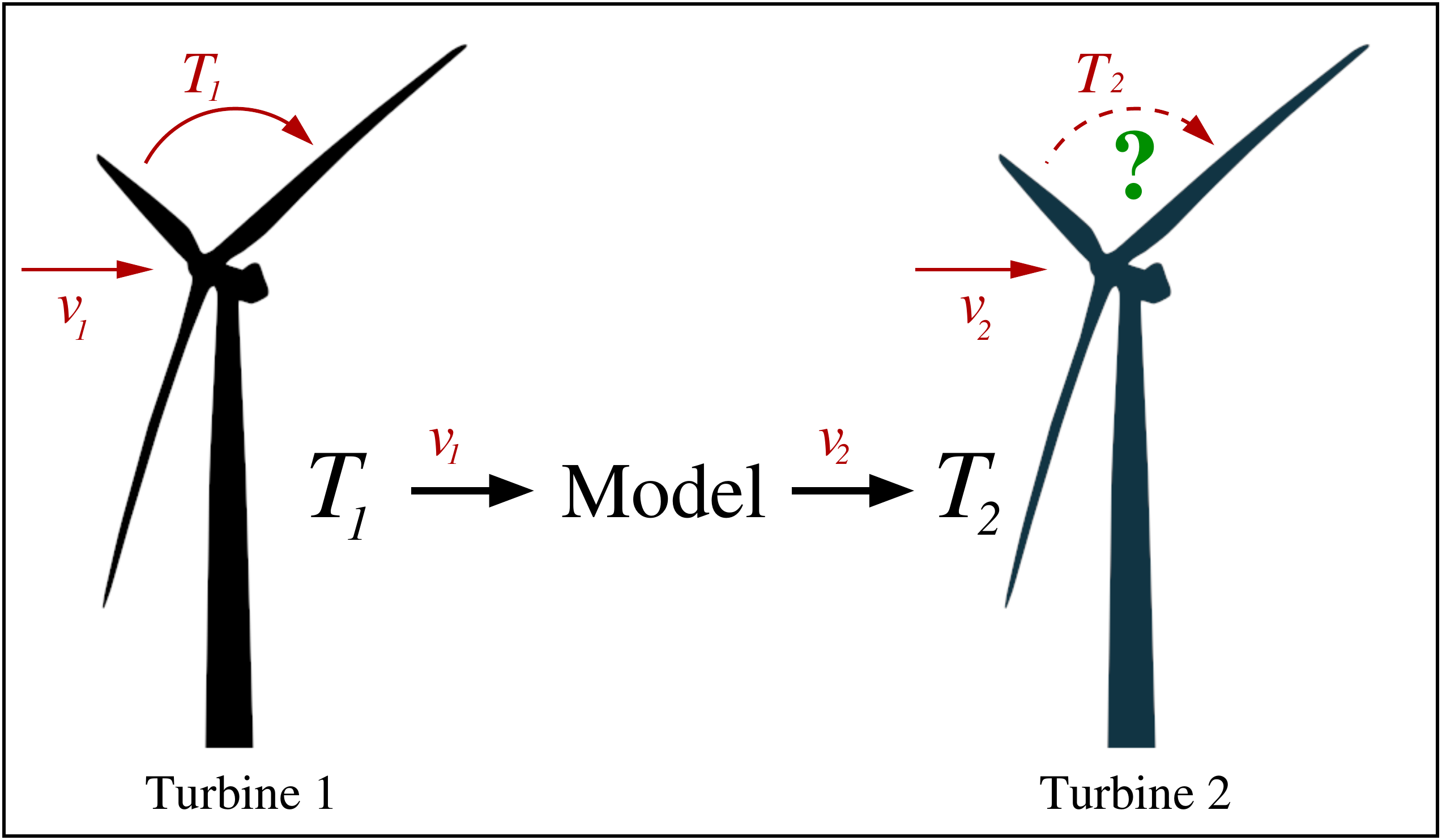}
\end{center}
\caption{\protect 
         Sketch of the proposed idea for estimating fatigue loads:
         one derives a stochastic model from the data series of the
         wind speed and torque measured at Turbine 1;
         using this model and the wind speed series measured at the 
         Turbine 2 one reconstructs the statistics of the
         torque at the Turbine 2; from the estimated torque
         series one estimates fatigue loads.
         This procedure can theoretically be applied to other
         properties one wants to access for monitoring and controlling
         wind turbines, avoiding to place measurement devices in
         all of them (see text).}
\label{fig01}
\end{figure}

Both these two different types of challenges in wind energy research
are in fact related with each other. 
An important example is that devices for measuring certain types of loads
are considerably expensive, and due to the unpredictable and
intermittent nature of the wind, one typically places one or 
more devices at each turbine, for preventing loss of accuracy 
in cumulative loads\cite{moriarty08,freudenreich08}.
The loads applied on one wind turbine are important to estimate,
within some accuracy, the life expectancy of a turbine or the maximal 
time-span between two inspections of its operating features. Moreover,
knowing how the applied loads evolve in time helps understanding the
fatigue behavior and critical situations that compromise the 
functionality of the turbine\cite{ragan2007}.
Therefore, establishing models for the intermittency of the loads
in single wind turbines would not only contribute for better understanding 
the energy production and the monitoring of wind turbines, but would also 
open the possibility for mitigating expensive procedures.

In this paper we propose a recent model for reconstructing the
increment statistics of the torque in single wind turbines
as a tool for estimating fatigue loads, not only of that wind
turbine but also of their neighboring ones.
As sketched in Fig.~\ref{fig01} we use the wind speed and torque
data series of one wind turbine (Turbine 1) to derive our stochastic
model that describes statistically the evolution of the torque 
subject to the observed wind speed.
We assume that the response of similar wind turbines to the
wind speed shall yield similar values of their properties,
such as the power, the blade bending moments or the torque on the 
main 
shaft. Therefore, we apply the derived stochastic model to another wind 
turbine (Turbine 2) where only the wind speed is 
measured on the nacelle. 
The outcome yields a series of torque estimates used for 
a fatigue analysis, including Markov matrices, rainflow 
counting statistics and load duration distribution, which 
are then compared with the fatigue analysis of the respective
torque measurements.
The overall result is that our stochastic model retrieves accurate
estimates of fatigue loads in any wind turbine within the same wind
farm.

We start in Sec.~\ref{sec:data} by describing the data analyzed and
in Sec.~\ref{sec:method} the method is described in further detail.
Our results for extracting the stochastic model are described in 
Sec.~\ref{sec:stochastic} and the fatigue analysis and fatigue estimate
are presented and discussed in Sec.~\ref{sec:fatigue}. 
Section \ref{sec:discuss} concludes the paper.

\section{Data and methodology}

\subsection{Wind speed and torque in Alpha Ventus}
\label{sec:data}

We analyze the set of measurements of the wind speed and torque
taken at two different wind turbines of Senvion in Alpha Ventus wind 
farm, namely turbines AV4 and AV5, during the full month of January 2013. 

In principle, our analysis could also be done for other types
of loads instead of the drive torque. However, in this article we focus 
on the fatigue on the drive train, since some of its components are known to
be prone to failure. In the future, we aim at extending our analysis to other 
turbine components.

\begin{figure}[t]
\centering
\includegraphics[width=0.85\textwidth]{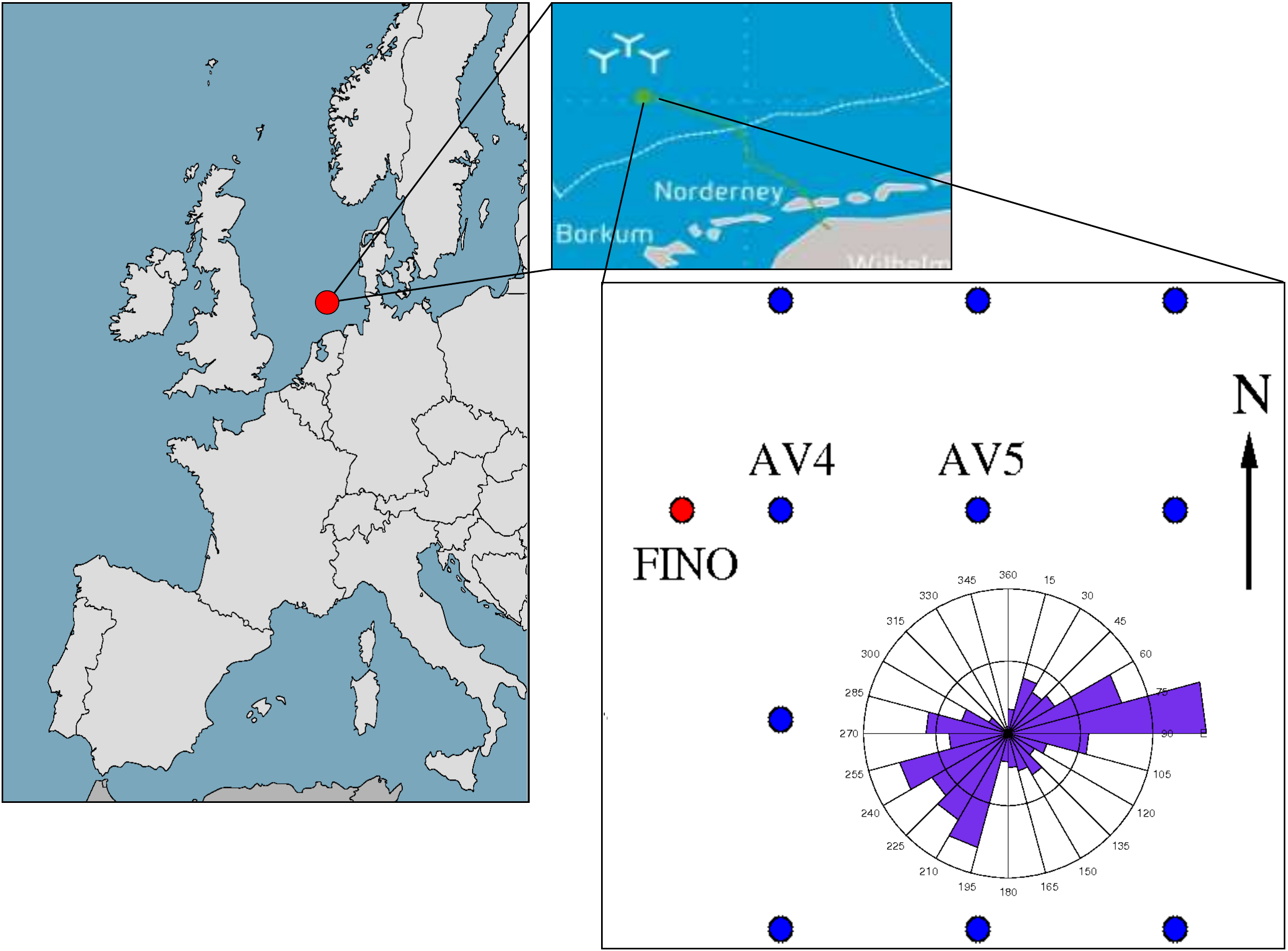}
\caption{\protect 
         Sketch of the Alpha Ventus wind farm, located in the North
         Sea, near the northeastern coast of Germany. The study here 
         described
         is based on measurements taken at the two wind turbines
         AV4 and AV5, manufactured by Senvion SE.
         In the period of January 2013, during which the 
           measurements analyzed in
           this paper were taken, the wind direction show a tendency
           towards East-northeast.}
\label{fig02}
\end{figure}

Figure \ref{fig02} shows a sketch of the Alpha Ventus wind 
farm, the first off-shore wind farm in Germany, located at Borkum 
West in the North Sea, $54.3^o$N-$6.5^o$W. 
As shown in the wind rose, during January 2013 the main wind
  direction lies between Northeast and East.

\begin{figure}[t]
\centering
\includegraphics[width=0.6\textwidth]{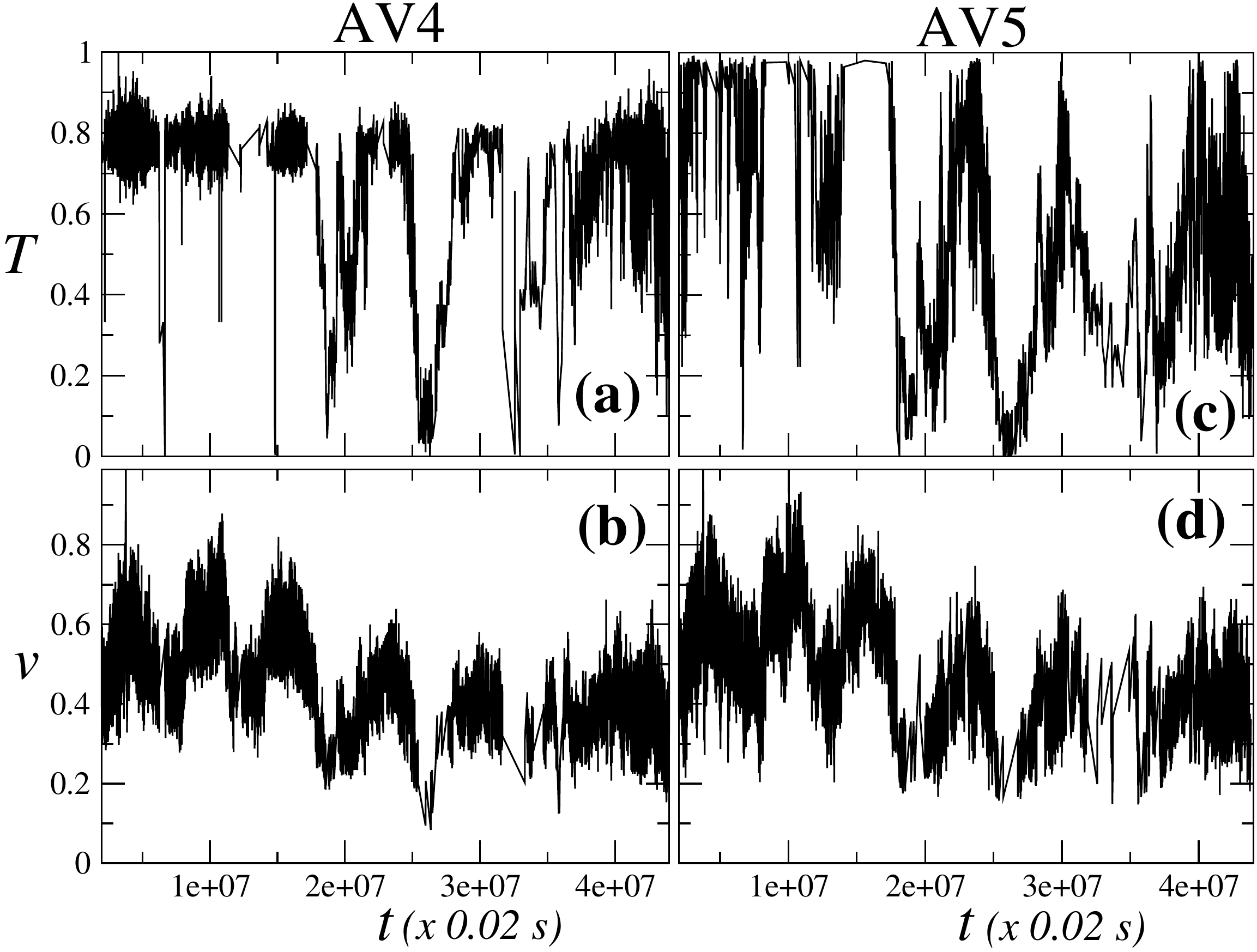}%
\includegraphics[width=0.4\textwidth]{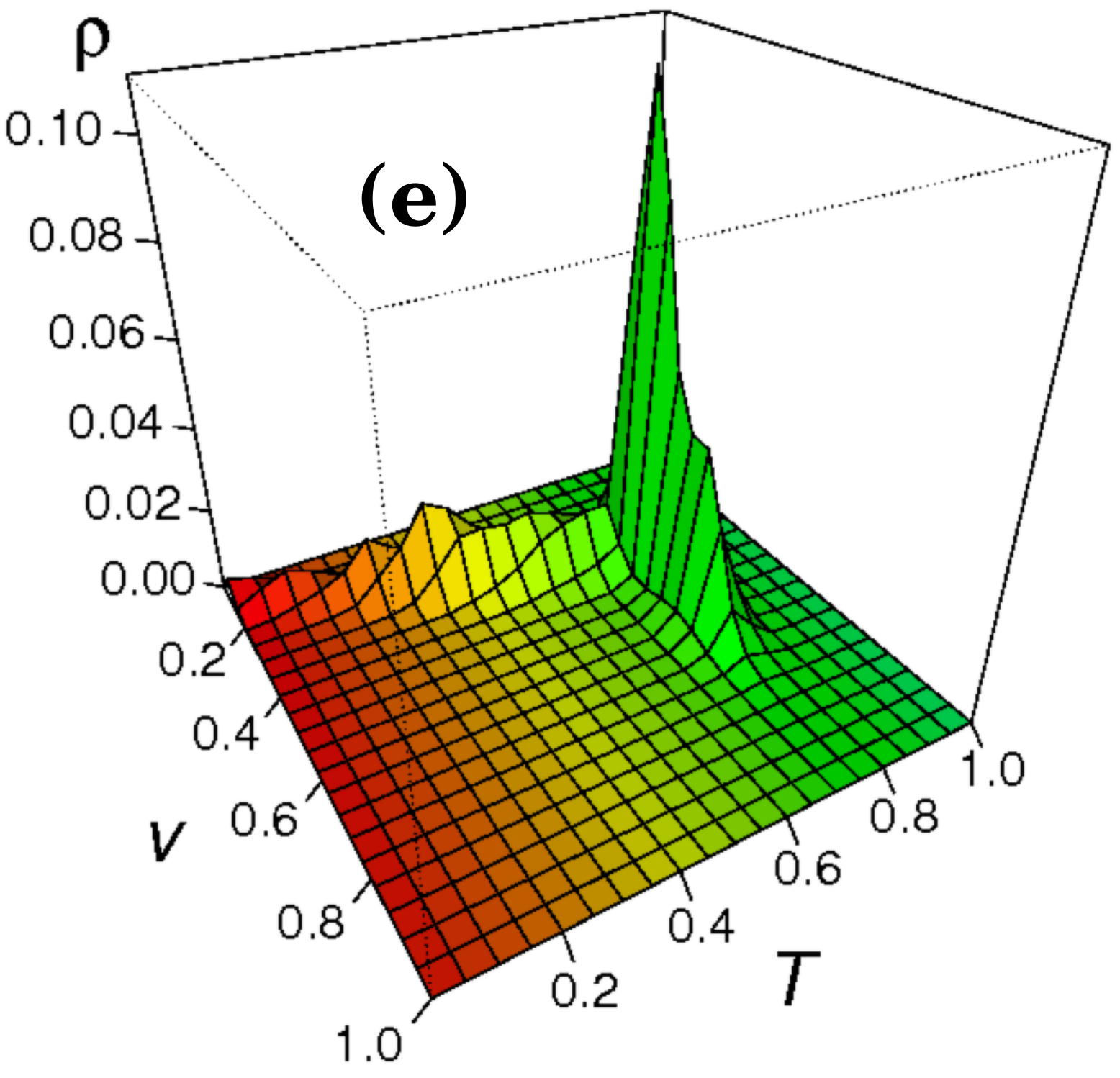}
\caption{\protect 
         The torque and wind speed measured at turbines AV4 and AV5
         from Senvion in Alpha Ventus wind farm.
         All data was masked through normalization to the
         largest values (see text).
         For constructing the stochastic model, one uses 
         {\bf (a)} the torque $T$ and 
         {\bf (b)} the wind speed $v$ at AV4, for estimating
         {\bf (c)} the torque at AV5 conditioned to its 
         {\bf (d)} wind speed.
        In {\bf (e)} we show the histogram on the observed  
         pair of torque and wind velocity measurements taken
         at turbine AV4. In both $T$ and $v$ $21$ bins were used
         to cover the full range of values $[0,1]$.
         This histogram will be of importance
         when addressing Fig.~\ref{fig05}.}
\label{fig03}
\end{figure}

The torque on the main shaft
$T$ is computed from the measurements of the power output 
$P$ and the rotor rotational speed $n$, 
assuming neither mechanical nor electrical losses,
\begin{equation}
T = \frac{30}{\pi}\frac{P}{n} 
\end{equation}
with $n$ in r.p.m., $P$ in W and $T$ in Nm.
The sampling rate of the power output, rotational speed 
and consequently
the torque is $50$ Hz and the sampling rate of the wind speed is $1$ Hz. 
Since we need to use the same sampling 
rate for all data series, we only consider torque measurements each
second, at instants for which a wind speed measurement also exists.
In total, for turbine AV4, we have 826745 data points.

Both the wind speed and torque measurements are shown in 
Fig.~\ref{fig03} and are according to the torque-speed curve 
known in the literature\cite{windenergyhandbook,philipposter,ourtorque}. 
As one sees, the torque at both wind turbines, in Figs.~\ref{fig03}a and 
\ref{fig03}c, shows periods of high power production (large values),
alternating with pronounced decreases and subsequent increase during which
large fluctuations occur. The largest values of the torque observed for turbine
AV4 (Fig.~\ref{fig03}a) are not so close to the maximum observed 
value as the values observed in the torque at AV5 (Fig.~\ref{fig03}c). 
As for the wind speed observed in both turbines (Figs.~\ref{fig03}b and 
\ref{fig03}d) it increases and decreases in an oscillating manner and 
simultaneously in both turbines, but showing no clear periodic pattern.
In Fig.~\ref{fig03}e one also plots the joint probability density 
function (PDF) for both the wind velocity and the torque, a plot that
will be of importance when interpreting the stochastic model introduced
below.

All data series were analyzed according to all confidential protocols and
were properly masked through the normalization by their highest value, being
here published with Senvion's permission.
Scientific conclusions are not affected by such data protection requirements.

Together with most of the properties measured at wind turbines, wind speed
and torque have typically fluctuations occurring at
different time-spans, defining series of increments, whose statistics 
describe the intermittency observed in the wind energy 
production\cite{muecke}. In the particular case of the torque, the
succession of torque increments are related with the so-called
random load cycles which are of importance for estimating
fatigue loads. 

From the physical point of view the fatigue load at one wind turbine
is roughly the time integral of load increments, or of some proper positive 
and monotonic function of the increments. 
Therefore, we consider the increments 
$\Delta T_{\tau}(t) = T(t+\tau)-T(t)$, taken with a fixed time-gap 
$\tau$.
As reported in Ref.~\cite{ourtorque}, for up to one hour or more,
the torque increment distributions are clearly non-Gaussian.
In Sec.~\ref{sec:method} we will describe a framework
which yields an evolution equation for the torque constrained to
the wind speed observed at the same turbine. With such a framework
we reconstruct not only the torque observed at that same wind
turbine, but also the torque observed in neighboring wind turbines
which we assume to respond similarly to the wind speed.

\subsection{The conditional Langevin model for turbine loads}
\label{sec:method}

The Langevin approach is a framework developed from the pioneer 
work by Peinke and Friedrich in 1997\cite{friedrich97}, which 
consists of a direct method for extracting the evolution 
equation of stochastic series of measurements.
Several applications were proposed and
developed, e.g.~in turbulence modeling, in medical EEG monitoring
and in stock markets. 
See Ref.~\cite{physrepreview} for a review.
In the context of wind energy, this framework has shown the ability
for predicting power curves of single wind 
turbines\cite{muecke,anahua2008,raischel2013} as well as
of equivalent power curves for entire wind farms\cite{milan2014}, and
also to properly reproduce the increment statistics of
power and torque in single wind turbines\cite{muecke,ourtorque}.

One considers a set of measurements $X(t)$ in time $t$ of one 
particular property $x$ evolving according to the so-called
Langevin equation defined by
\begin{equation}
\frac{dx}{dt}=D^{(1)}(x)+\sqrt{D^{(2)}(x)}\Gamma_t ,
\label{LangVect}
\end{equation}
where $\Gamma_t$ is a Gaussian $\delta$-correlated white noise,
i.e.~$\langle \Gamma(t)\rangle = 0$ and 
$\langle \Gamma(t)\Gamma(t')\rangle = 2\delta_{ij}\delta(t-t')$. 

The first term in the right hand-side incorporates the deterministic 
contributions in the process, yielding the drift function $D^{(1)}$,
while the second terms models the diffusion, i.e.~the total stochastic 
contributions accounting for the stochastic fluctuations, which are 
incorporated by function $D^{(2)}$.
The constant in $\delta$-correlation and the square root in 
the Langevin equation are usually chosen for convenience.
Details can be found in Ref.~\cite{physrepreview}.


Since the drift and diffusion function have a physical interpretation,
one could apply the model in Eq.~(\ref{LangVect}) to a particular system
and define {\it ad hoc} the functional shape of both functions from
physical reasoning. 
In several cases however such approach is not convenient.
For instance, when considering properties which result from the 
interaction between a turbulent system, such as the atmosphere,
and a complex technical device, such as the drive train in a wind turbine.
In those cases, it is hard to derive drift and diffusion from first 
principles and the one natural alternative to get information about 
their functional form is data analysis.
In this work we deal, therefore, with the
inverse problem: 
having a set of measurements is it possible to derive the drift
and diffusion functions in order to reproduce the statistics of
that set of measurements from simple integration of Eq.~(\ref{LangVect})?

The short answer is yes, it is possible.
The long answer has two main steps.

In the first step,
one needs to test the data to ascertain if it can be taken as a
Markovian series, i.e. if there is a time interval $t_{\ell}$,
so-called the Markov-Einstein length, 
for which in the succession of measurements 
the next value only depends on the present one and is independent of 
the values previous to it. 
In other words, one must test if 
\begin{equation}
\rho(X(t+t_{\ell})\vert X(t), X(t-t_{\ell}), X(t-2t_{\ell}), \dots) =
\rho(X(t+t_{\ell})\vert X(t))
\label{condMarkov}
\end{equation}
holds,
with $\rho(x\vert y)$ representing the conditional probability density 
function (PDF) for observing $x$ having observed $y$. This PDF
can be extracted from histograms of the data set.

There are simple standard ways to perform this test\cite{physrepreview},
which typically compare three-point statistics such as 
$\rho(X(t+t_{\ell})\vert X(t), X(t-t_{\ell}))$ conditioned to a 
fixed value $X(t-t_{\ell})=X^{\ast}$, with
the corresponding two-point statistics $\rho(X(t+t_{\ell})\vert X(t))$ 
and ascertain for which $t_{\ell}$ one obtains the best overlap, between
both PDFs.

\begin{figure}[t]
\centering
\includegraphics[width=0.8\textwidth]{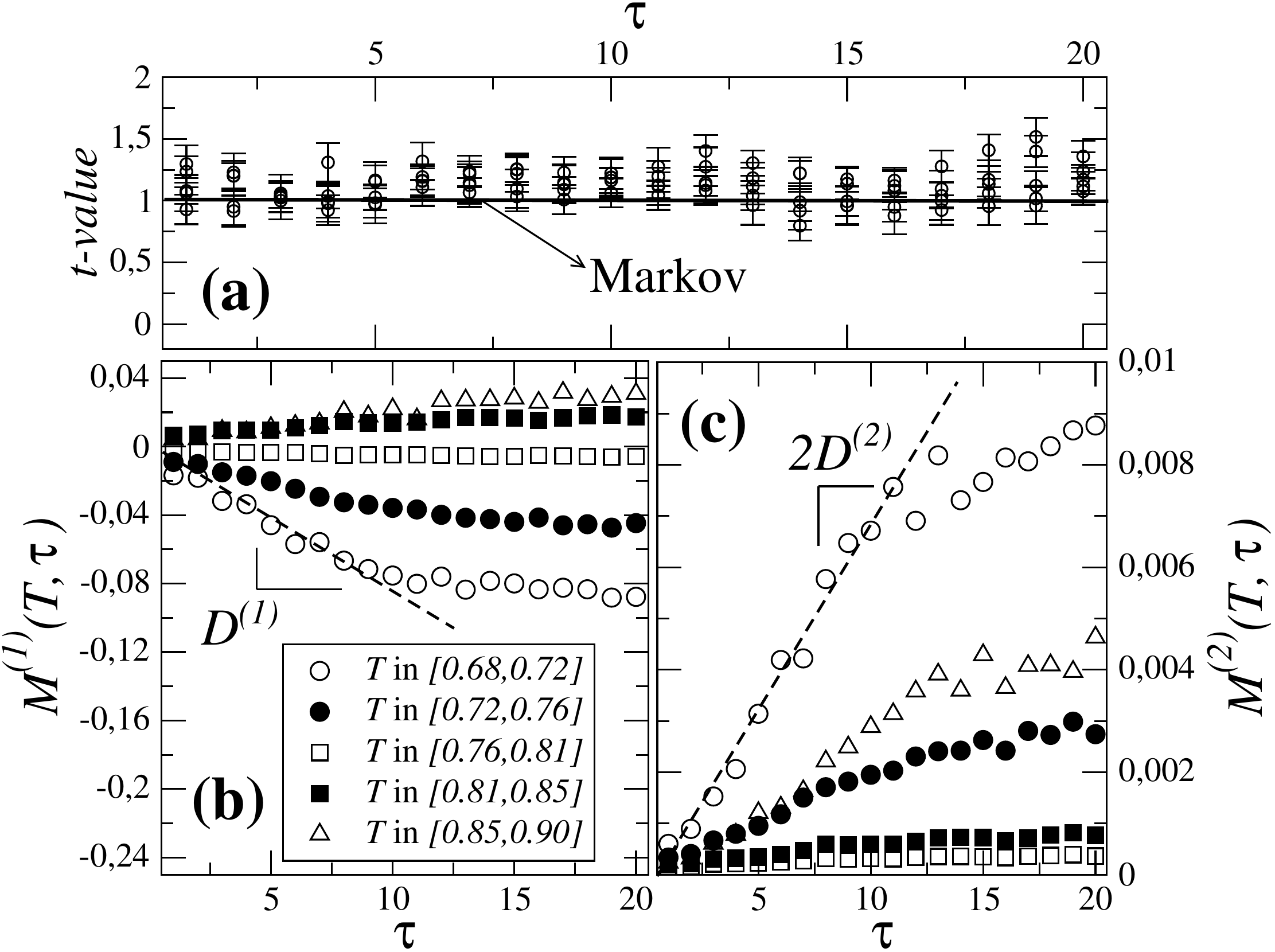}
\caption{\protect 
         The Markov-Wilcoxon test retrieves a normalized
              $t$-value which taking the value $1$ (solid line)
              indicates perfect matching between the two condition 
              probability distributions. 
              {\bf (a)} In the load series conditioned to wind 
              velocity values the Markov condition is not perfectly
              fulfilled, but the method is still able to reproduce
              accurate estimates (see text).
              For that estimate one starts by computing 
              {\bf (b)} the first and
              {\bf (c)} the second conditional moments as defined 
              in Eqs.~(\ref{condmoments}). As indicated with dashed 
              lines the slope observed yields the values for the
              drift and diffusion equation 
              (see Eq.~(\ref{DefCoefKM})).}
\label{fig04}
\end{figure}

Though, when the measurements obey this Markov condition the next step
can be carried out, one advantage of this Langevin framework
is that it also works in some cases where the Markov test fails.
One of such cases is the presence of measurement 
noise\cite{boettcher2006,lind2010}, which opens a broad panoply of
different practical situations where this framework is applicable.
In the present case, though no measurement noise seems to
be present, as reported below, the Markov condition is not perfectly
fulfilled as shown in Fig.~\ref{fig04}a.
Still, important to stress here, it is possible to accurately estimate
fatigue loads using such a framework.

To test the Markov condition we compute a $t$-value as shown in
Fig.~\ref{fig04}a.
The Wilcoxon test is a statistical test used to test if Eq.~(\ref{condMarkov}) holds.
One assumes to have two different series of values from independent
random variables and tests whether the variables are identically
distributed or not.
The model retrieves a $t$-value, which when equal to unit indicates
perfect matching between both distributions in Eq.~(\ref{condMarkov}).
As we see from Fig.~\ref{fig04}a, though the observed $t$-values lie
near the perfect fit value, they can not always be taken as being one
within numerical precision. Still, as we explain below, such fact does 
not prevent us for accurately estimate fatigue loads.

For the second step,
one needs to derive both functions, $D^{(1)}$ and $D^{(2)}$,
that fully define Eq.~(\ref{LangVect}). 
The derivation is performed through the computation of the
corresponding first and second conditional moments\cite{ourtorque},
respectively
\begin{subequations}
\begin{eqnarray} 
M^{(1)}(x,\tau) &=& \left\langle X(t+\tau)-X(t)\right\rangle |_{X(t)=x} \\ 
M^{(2)}(x,\tau) &=& \left\langle (X(t+\tau)-X(t))^2 \right\rangle |_{X(t)=x} 
\end{eqnarray}
\label{condmoments}
\end{subequations} 
where $\langle \cdot \rangle |_{X(t)=x}$ indicates the average
over the full time series, whenever $X(t)$ takes the value $x$.

Figures \ref{fig04}b and \ref{fig04}c show the first and second
conditional moments for five different values of the torque at AV4.
By taking the slope of the linear regression for each conditional
moment yields the corresponding coefficient $D^{(1)}$ and $D^{(2)}$.
Indeed, it can be shown\cite{risken} that drift and diffusion functions in 
Eq.~(\ref{LangVect}) are, apart from a
multiplicative constant ($1/k!$), the derivative with respect to the 
time-gap $\tau$ of the first and second conditional moments 
respectively, which for arbitrary order $k$ is defined as
\begin{equation}
D^{(k)}(x)=\lim_{\tau\rightarrow0}\frac{1}{k!}\frac{M^{(k)}(x,\tau)}{\tau} .
\label{DefCoefKM}
\end{equation}
Therefore, drift ($k=1$) and diffusion ($k=2$) can be directly extracted 
from the data sets.
For accuracy purposes a correction is introduced for the diffusion
function\cite{gottschall2008}, where instead of the second condition
moment $M^{(2)}(x,\tau)$ one considers
\begin{equation}
M^{(2)}_{cor}(x,\tau) = M^{(2)}(x,\tau)-[M^{(1)}(x,\tau)]^2 .
\end{equation}

Within a sufficiently low range of $\tau$ values,
typically between five and ten time-steps of the set of measurements, 
the conditional moments depend linearly on $\tau$. 
Thus, for each value $x$, both $D^{(1)}(x)$ and $D^{(2)}(x)$ are given 
by the slope of the linear interpolation of the corresponding conditional 
moments.

It is worth mentioning that, as known in previous 
studies\cite{boettcher2006,lind2010}, the existence of an offset
in the conditional moments indicates the presence of measurement
noise in the data. As shown in Figs.~\ref{fig04}b and \ref{fig04}c
the linear interpolations cross the origin and no offset seems
to exist. Therefore one can say that measurement noise is not
significant in data sets analyzed in this paper.

On top of these two steps, one important assumption must be added: the 
set of measurements must be stationary. 
This is of course {\it not} the case for torque series at one wind turbine.
To overcome this shortcoming, it was recently proposed 
that\cite{milanpriv,ourtorque}
one should consider the succession of measurements of the torque
corresponding to a sufficiently narrow range of speed values.
Considering only this subset of torque values governed by what we
named the conditioned Langevin equation\cite{ourtorque}, one observes that the 
statistical moments of the torque values distribution are approximately 
constant, depending only on the wind speed (not shown). The conditioned 
Langevin equation used here as model is defined as
\begin{equation}
\frac{dT}{dt}=D^{(1)}(T,v)+\sqrt{D^{(2)}(T,v)}\Gamma_t,
\label{eq:langevinT}
\end{equation}
where, for our purposes, $T$ represents the torque at the wind turbine and 
$v$ is the wind speed.

\begin{figure}[t]
\centering
\includegraphics[width=0.43\textwidth]{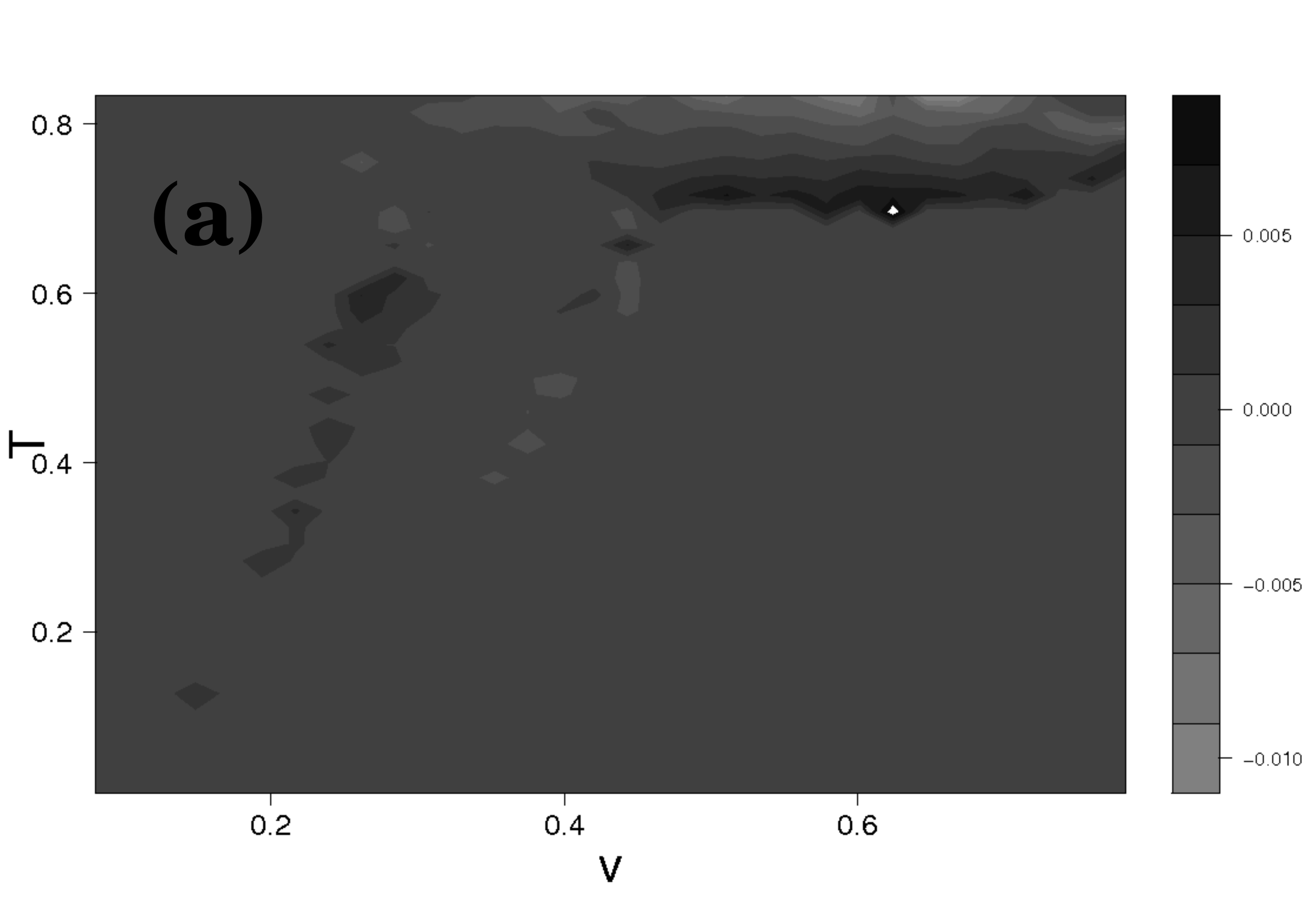}%
\includegraphics[width=0.43\textwidth]{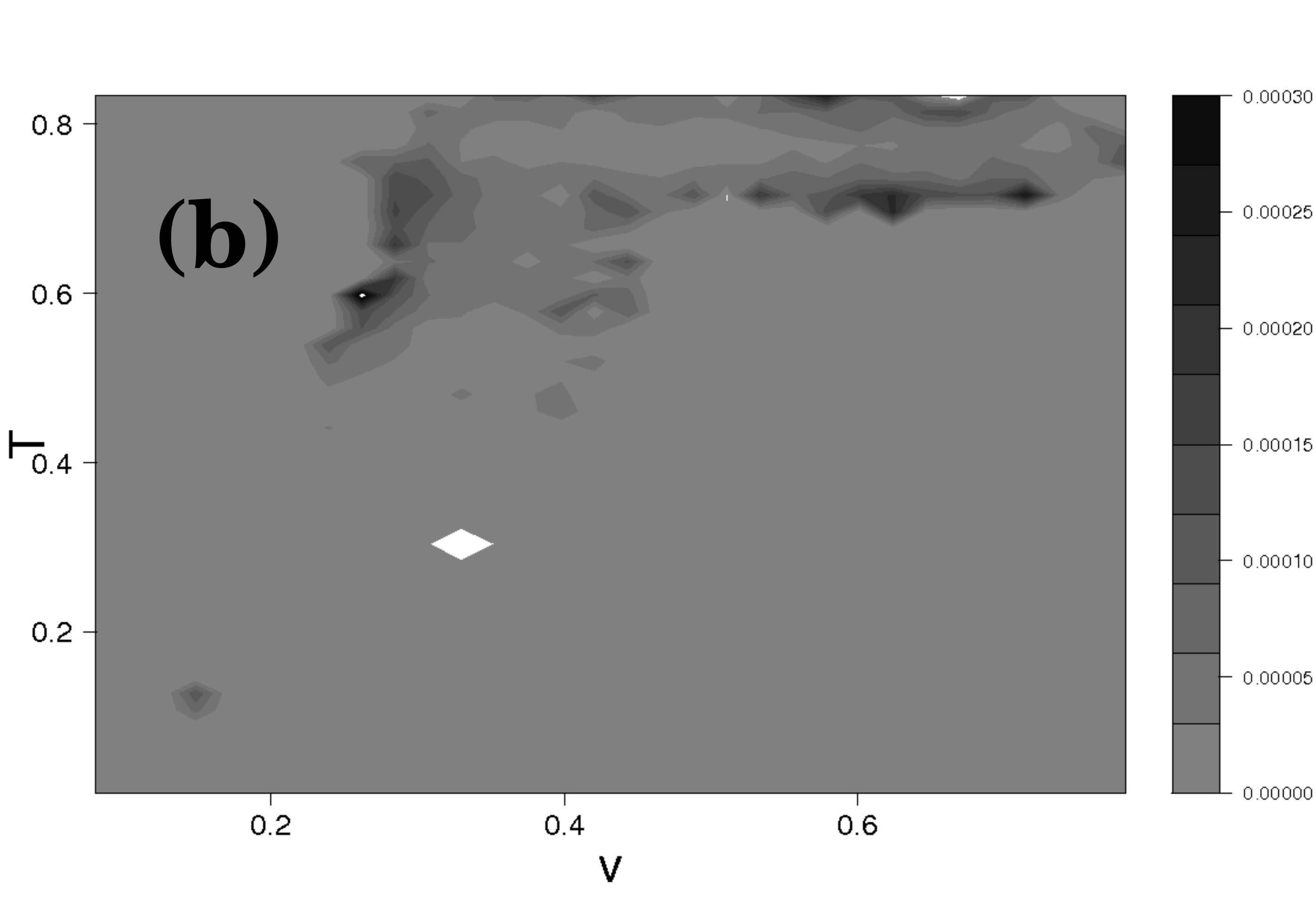}
\includegraphics[width=0.45\textwidth]{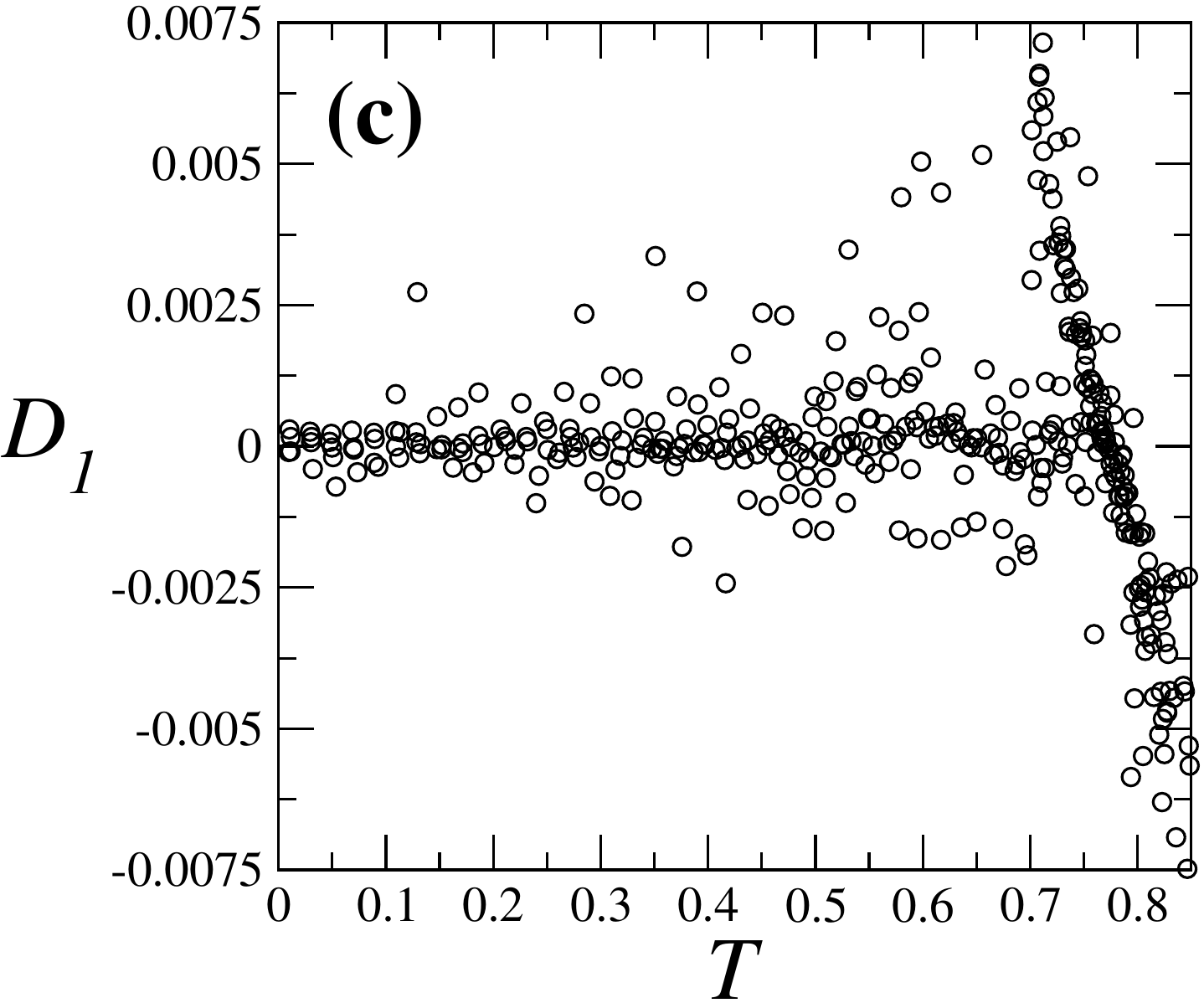}%
\includegraphics[width=0.41\textwidth]{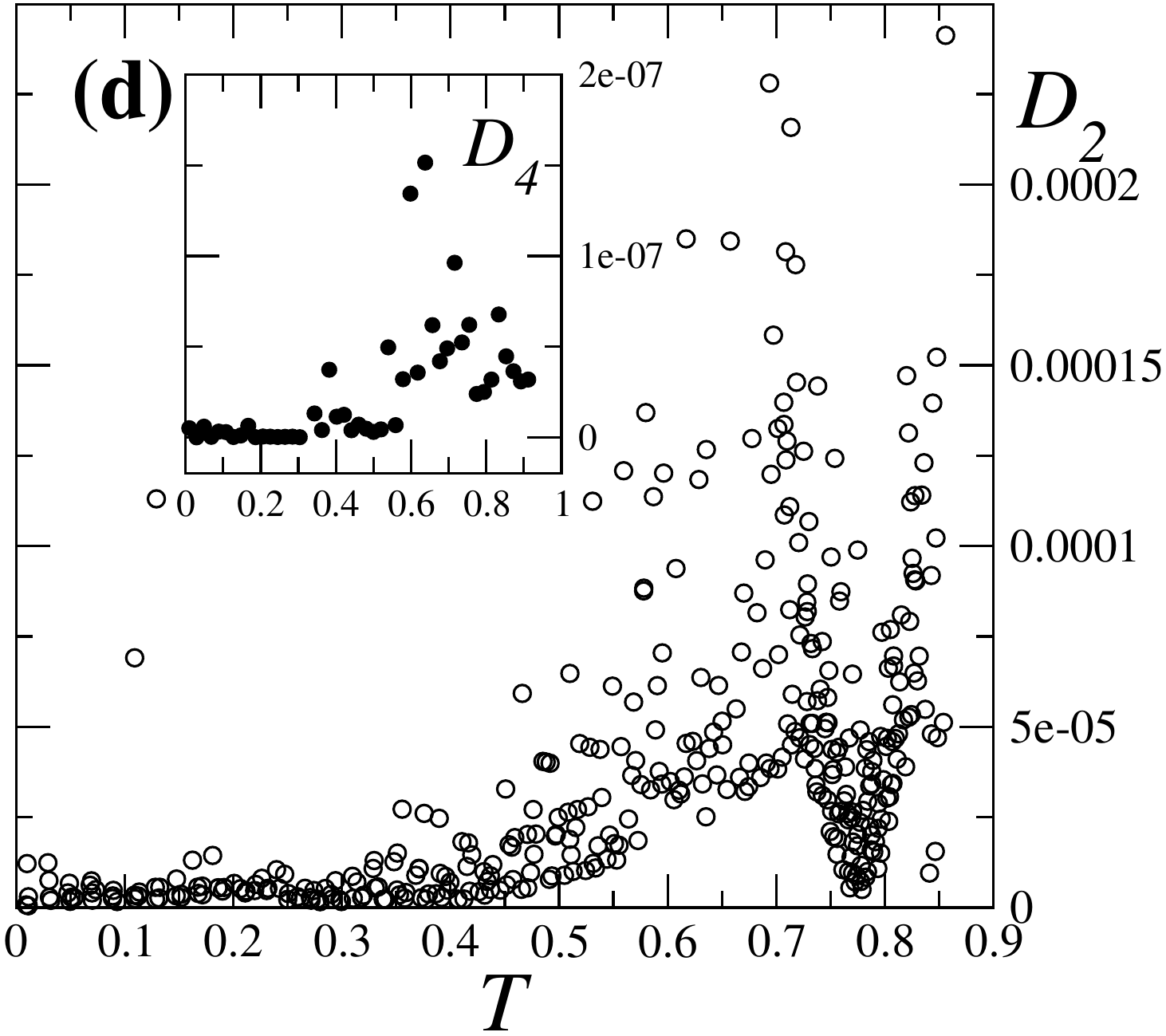}
\caption{\protect 
         Numerical result for 
         {\bf (a)} the drift $D^{(1)}(T,v)$ and 
         {\bf (b)} the diffusion $D^{(2)}(T,v)$ in the Langevin equation 
         (\ref{eq:langevinT}) from which the time series of the torque 
         is reconstructed. 
         In 
         {\bf (c)} and 
         {\bf (d)} one plots the corresponding
         integration over the wind speed.
         The numerical values of both functions are the ones used
         for predicting the torque series in wind turbine AV5. In the inset
         one sees the lower values of $D^{(4)}$ which enables us to
         only consider drift and diffusion (see text). 
         Here, only the range $[0,0.85]$ is plotted; as shown in
         Fig.~\ref{fig03}e beyond this range no significant statistics
         is observed. 
         Both $D^{(1)}$ and $D^{(2)}$ were computed for $21$ bins in the 
         torque, covering only the observed range of values at each particular
         velocity bin ($21$ bins).}
\label{fig05}
\end{figure}

For such extension of the Langevin approach, now conditioned to the
values the wind velocity, all the procedure described above is applied
separately for value of the wind velocity. 
This means that, having the full series of the torque at AV4, one now
filters out all values except those observed together with a given
wind velocity $v=v^{\ast}$ and computes, one for them, the 
conditional moments:
\begin{subequations}
\begin{eqnarray}
M^{(1)}(T,\tau) &=& \left\langle T(t+\tau)-T(t)\right\rangle
|_{T(t)=T,v(t)=v^{\ast}} , \\
M^{(2)}(T,\tau) &=& \left\langle (T(t+\tau)-T(t))^2 \right\rangle
|_{T(t)=T, v(t)=v^{\ast}} .
\end{eqnarray}
\label{condmoments2}
\end{subequations}
Similarly, the Markov test was also performed for each wind velocity
separately, as described above.

Having derived both drift and diffusion functions a final test must
be performed. 
Generically, one can compute from Eqs.~(\ref{DefCoefKM}) the
coefficients $D^{(k)}(x)$ of an arbitrary order $k$.
However, according to Pawula's theorem\cite{risken}, in the case
that $D^{(4)}$ is zero or, for practical purposes, it is negligible in
comparison with the drift and the diffusion, all coefficients of order 
$k\geqslant 3$ are also negligible.

Figure \ref{fig05}a and \ref{fig05}b show the drift and diffusion
respectively as a function of both the torque and the wind velocity.
The projection along the velocity is shown in Figs.~\ref{fig05}c and
\ref{fig05}d where one identifies a transition approximately at 
$T=0.75$ where the drift depends linearly on the torque, while the
diffusion depends quadratically on the torque.

From the plots \ref{fig05}a and \ref{fig05}b one sees that the value
$T=0.75$ is first attained for wind velocities around $v=0.4$. These
two values also separate two different regions when observing the
joint PDF (see Fig.~\ref{fig03}e).
Above this value the controller at a wind turbine starts operating.
Below that velocity the drift is almost absent and diffusion decays
with the magnitude of the velocity.
Considering Eq.~\ref{eq:langevinT}, a closer look into the plots in
Fig.~\ref{fig05} enables one to physically interpret $D^{(1)}$ and 
$D^{(2)}$ in a stochastic model for the torque conditioned to the
wind velocity.

In general\cite{risken}, 
the drift measures the force driving a specific property
to increase or decrease in time, while the amplitude in the diffusion
coefficient measures the amplitude of a fluctuation around that driven
motion.
Having such physical interpretation in mind and looking again to the 
plots in Fig.~\ref{fig05} one can conclude that the controller starts
operating around $v=0.4$. Indeed, for this value the drift coefficient
$D^{(1)}$ shows a linear dependence on the torque with a negative 
slope: when the torque increases the controller acts in order to decrease 
it and vice-versa. Simultaneously, diffusion depends
quadratically on $T$.
Moreover, 
below the velocity threshold for controller operation the drift almost
vanishes with a diffusion that decreases with the magnitude of the
wind velocity: no force is driving the torque for this range of wind 
velocities and the magnitude of torque fluctuations increases with
the wind velocity.

\section{The stochastic model for fatigue loads}

Using the Langevin framework described in the previous section,
we next apply it to wind turbine AV4 in order to extract a
stochastic model for the torque, that is assumed to hold for any 
Alpha Ventus turbine of the same model. Then, we 
use the set of wind speed measurements at a different wind turbine,
namely AV5, for generating the increments of the torque at
this second turbine and compare it with the torque measurements.
Finally we show that the estimate of the fatigue loads at wind
turbine AV5 are statistically the same for the empirical torque
measurements and for the reconstructed series.

\subsection{Reconstruction of non-stationary time-series}
\label{sec:stochastic}

In a previous paper\cite{ourtorque} we have shown that the anomalous wind
statistics are responsible for the intermittent time evolution of the 
load in single wind turbines, promoting additional fatigue of the turbine 
itself.
The approach used there follows from the method proposed in 
Ref.~\cite{patrickprl,muecke} to the power output of 
single turbines.

We now use the model introduced in our previous work, with the 
drift and diffusion functions, $D^{(1)}$ and $D^{(2)}$ in Fig.~\ref{fig05},
extracted according to the framework described above for turbine
AV4, and apply it to
generate the torque in a neighboring wind turbine, namely AV5.

\begin{figure}[t]
\centering
\includegraphics[width=0.9\textwidth]{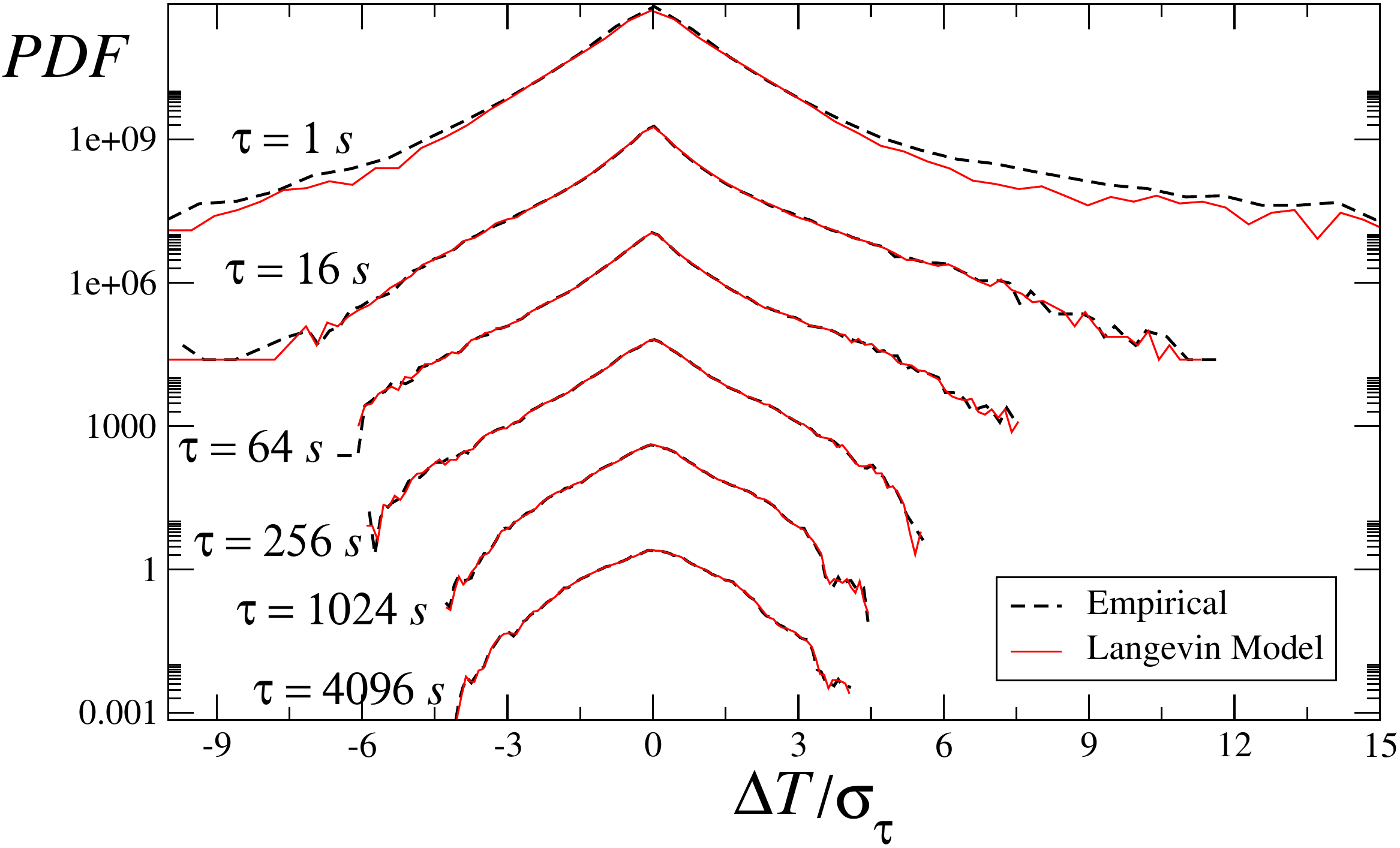}
\caption{\protect 
         Reconstruction of 
         the probability density function of the increments 
         (fluctuations) of the torque at turbine AV5.
         Time is in seconds, and the torque increments is in units of
         corresponding standard deviations ($\sigma_{\tau}$).}
\label{fig06}
\end{figure}

The wind speed measurements taken at AV5 are used as input data,
and initialize the differential evolution equation, Eq.~(\ref{eq:langevinT}),
with the first measurement of the torque at this same wind turbine.
The reconstruction of the increment statistics is plotted in Fig.~\ref{fig06}
with solid lines together with the empirical distributions in
dashed lines. While for low values of the time-span $\tau$ the extreme
fluctuations are not completely well reproduced,
above a time window larger than 
a few seconds the fit between measurements and model is very good.

In the inset of Fig.~\ref{fig05}b one sees that the typical values of
$D^{(4)}$ are of the order of $1$ to $2\ \%$ of $D^{(2)}$ values. 
As explained above, such low
values corroborate the applicability of the method to these data sets.
\begin{figure}[t]
\centering
\includegraphics[width=0.45\textwidth]{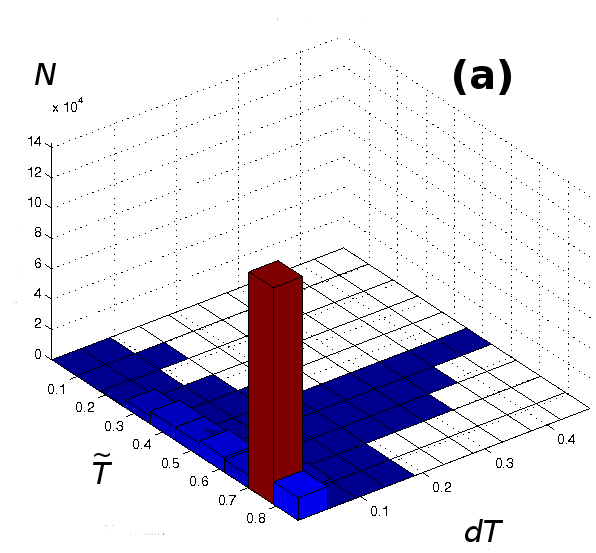}%
\includegraphics[width=0.45\textwidth]{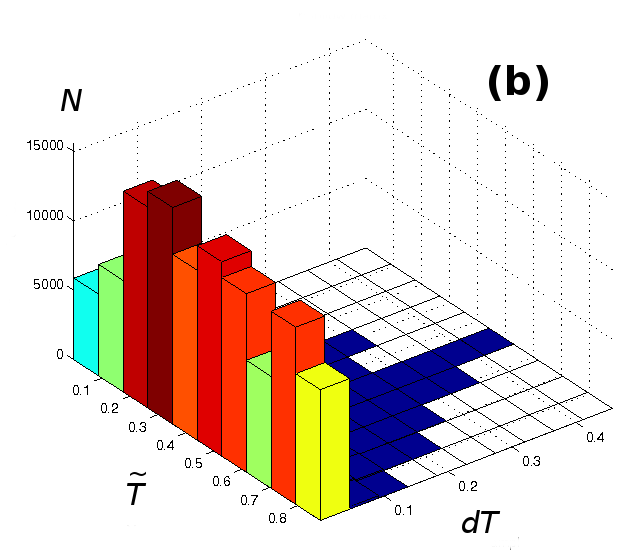}
\includegraphics[width=0.45\textwidth]{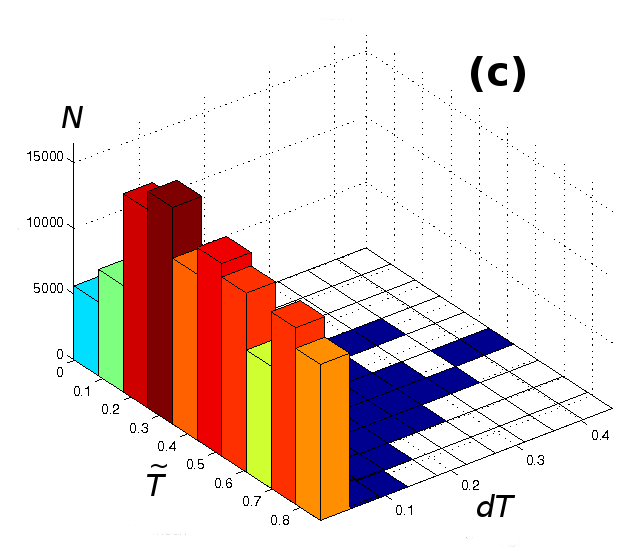}
\caption{\protect 
         Plot of the Markov matrices, i.e.~histograms of the
         number of cycles $N$ as a function of the mean value of
         every load cycle, $\tilde{T}$, and
         the cycle amplitude itself $\Delta T$ for
         {\bf (a)} the torque measurements of AV4,
         {\bf (b)} the torque measurements of AV5, and
         {\bf (c)} the torque estimates of AV5.}
\label{fig07}
\end{figure}

For practical purposes, one should briefly stress that the 
model here described is suitable for cases where the 
sampling frequency of the wind speed and torque data is high enough.
A sparse sampling of the wind speeds leads to a poor increment
statistic and consequently is not accurate enough for estimating fatigue 
loads.

Having a proper way for reconstructing the increment statistics, one
is now in position to use the reconstructed series of loads for estimating
the fatigue at AV5 and compare it with the empirical fatigue analysis. 
\begin{figure}[htb]
\centering
\includegraphics[width=0.85\textwidth]{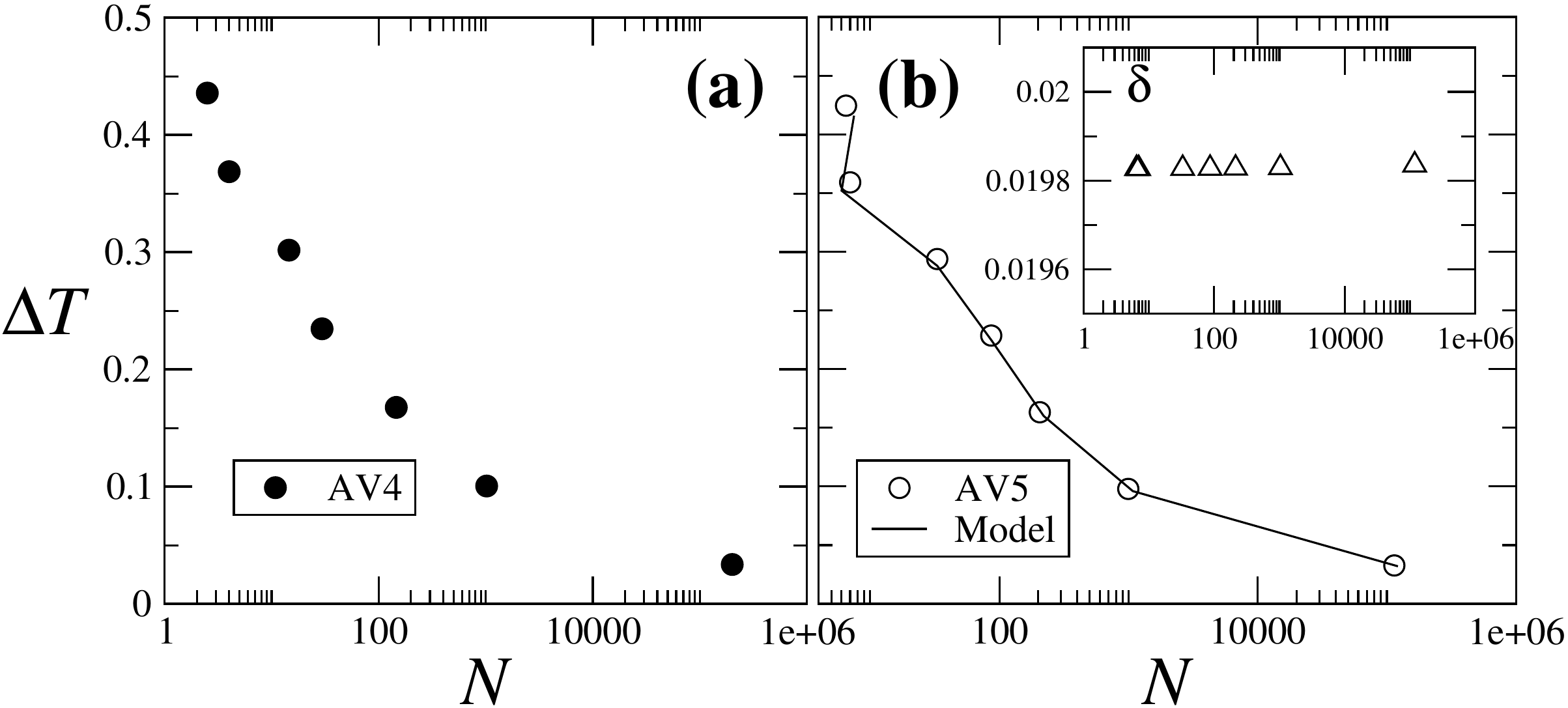}%
\caption{\protect 
         Rainflow counting (RFC) of the load (torque) series at
         {\bf (a)} AV4 (the data set used for deriving the drift and
         diffusion of the torque model plotted in Fig.~\ref{fig05})
         and {\bf (b)} AV5. 
         While the measurements at AV4 were used for deriving the 
         Langevin model, for
         AV5 one uses the derived model for reconstructing the data
         and compare the estimated loads with the observed ones.
         In the inset we show the relative deviation $\eta$ from 
         Eq.~(\ref{delta}) of the estimate with respect
         to the experimental results.}
\label{fig08}
\end{figure}

\begin{figure}[htb]
\centering
\includegraphics[width=0.85\textwidth]{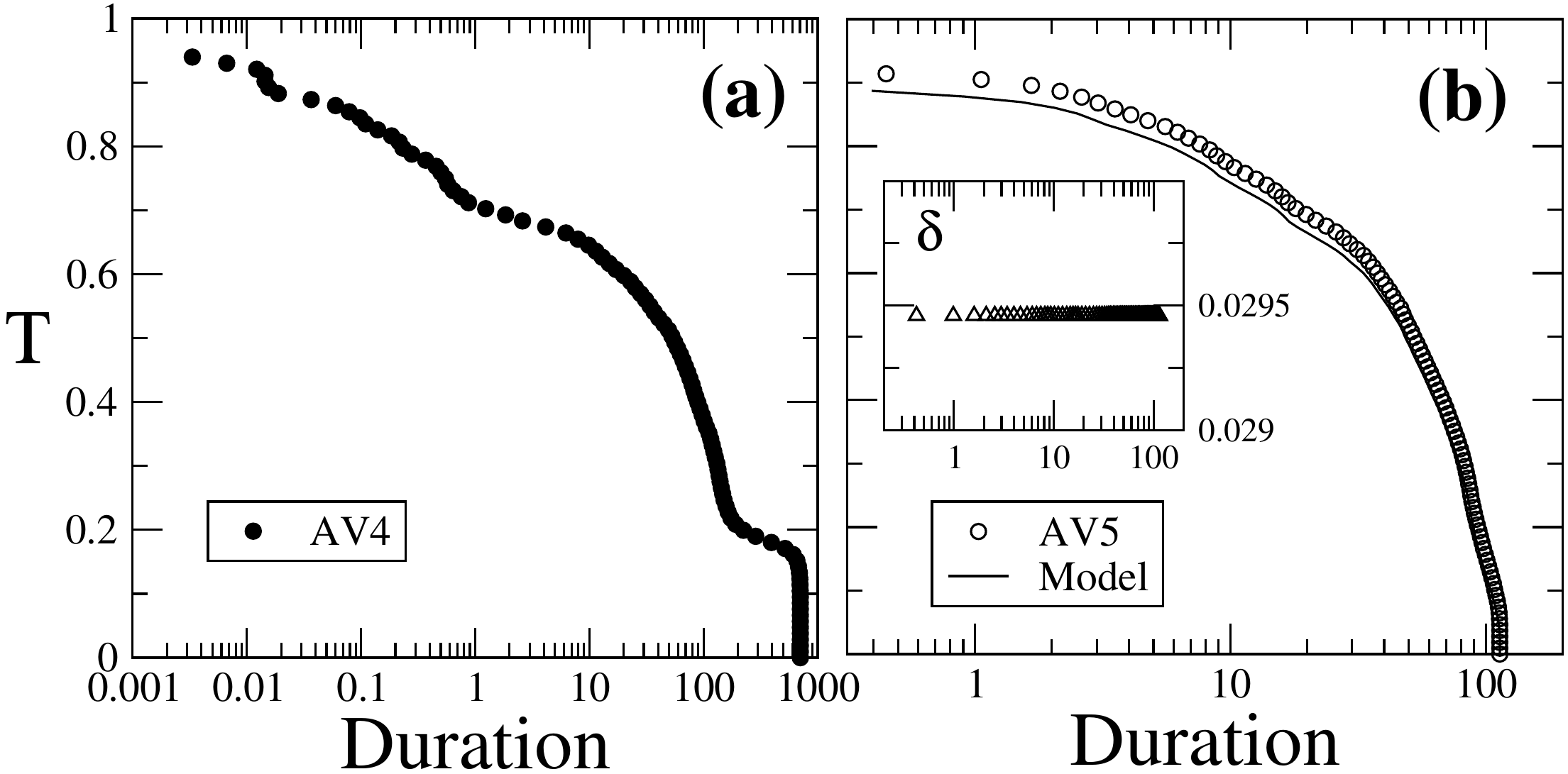}
\caption{\protect 
         Load duration distribution (LDD) observed at the load 
         series at
         {\bf (a)} AV4 and {\bf (b)} AV5. While for AV4 the
         empirical data was used for deriving the model, for
         AV5 one uses the derived model for reconstructing the data
         and compare the estimated loads with the observed ones.
         In the inset we show the relative deviation $\eta$ shown in 
         Eq.~(\ref{delta}) of the estimate with respect
         to the results for the measurements.}
\label{fig09}
\end{figure}



\subsection{Fatigue analysis and estimate}
\label{sec:fatigue}

To test the capability for the Langevin framework to predict fatigue
loads in a wind farm, we next present a comparative fatigue analysis 
between both the empirical data sets at AV4 and AV5 and the reconstructed
data set for AV5 only.
This comparative analysis is based on three standard tools, namely 
Markov matrices of the number of cycles, the rainflow counting 
procedure and the load duration distribution. All these tools are based 
in the concept of load cycle which is the 
observed load fluctuation 
between two successive local 
minima or maxima of the load.

The rainflow-counting method was proposed in the late sixties by
Endo and Matsuishi\cite{endo} and became the most standard algorithm 
for cycle-counting during the last decades, due to its ability for
accurately predicting the fatigue loads in a sequence of 
non-constant cycles.
In short, this method aims to count the  number of local maxima 
and minima weighting each one by its increment fluctuation, defined
mathematically by Rychlik in Ref.~\cite{rychlik} as follows.

Having a time-series $X(t)$ for $0<t<T$, let $N(t)$ be the number of counted
cycles  with maximum $M_k$ and minimum $m_k<M_k$ and let $N_T(u,v)$ be
the count distribution yielding the number of cycles counted in $X(t)$
such that $m_k<v\ge u<M_k$. The rainflow counting method assumes that
the count intensity $\mu(u,v)$ is the asymptotic behavior  of the expected value
of the count distribution per time unit:
\begin{equation}
\mu(u,v) = lim_{T\to \infty} \frac{E[N_T(u,v)]}{T}
\label{mu}
\end{equation}

In the context of fatigue analysis, 
the so-called Markov matrix
\footnote{The term ``Markov matrix'' is used here for the plots shown
in Fig.~\ref{fig07} and is standard in the context of engineering
for fatigue analysis. It should not be confused with the terms
``Markov process'' and ``Markov condition'' used in previous sections,
standard in the field of Statistical Physics, with which other different
matrix - e.g.~transition matrices - may be associated.} retrieves the number
of load cycles of a given amplitude $\Delta T$ as a function of the
mean value of the observed load during each load cycle 
$\tilde{T}$. In Fig.~\ref{fig07}a and \ref{fig07}b we plot 
the Markov matrix of the observed load in AV4 and AV5 respectively.
In both cases most of the cycles have an amplitude smaller than $0.5$,
in units of the maximum value of the load.
While at AV4 such amplitudes are observed for large loads,
around $0.7$, at AV5 such amplitudes are also observed in the remaining
range of values, being predominant in the lower load region, around $0.3$.
The reconstructed load series for AV5, shown in Fig.~\ref{fig07}c, however 
retrieves a good estimate of the full range of values, still with same 
deviation for the highest load values.

The difference between the observed Markov matrices for AV4 and AV5,
with AV5 showing a broader range of torque strengths can be explained by
considering the location of each turbine with respect to the main wind
directions.
As sketched in Fig.~\ref{fig02}, AV4 lies westerly from AV5, which lies
in the middle of four neighboring wind turbines. Further, the main
wind directions are southwest and east-northeast. Thus, the wake
effects on AV5 should be stronger than on AV4 and consequently the
optimal operation is not as frequent.

Similarly to what we did with Markov matrices we compute the 
rainflow counting (RFC) for AV4 (bullets in Fig.~\ref{fig08}a) and
AV5 (circles in Fig.~\ref{fig08}b). While the RFC spectra of AV4 and AV5 
are qualitatively different, the Langevin model applied to AV5 retrieves 
a correct estimation of the rainflow spectra. In the inset of 
Fig.~\ref{fig08}b one sees the relative error 
\begin{equation}
\eta=\frac{A_{r}-A_{e}}{A_{r}} ,
\label{delta}
\end{equation}
where $A_r$ and $A_e$ are the amplitudes observed for the real and
estimated values. The relative errors of the rainflow counting for
the estimates is less than $2\ \%$.

Finally, the load duration distribution (LDD) 
shows the amount of time a load with a given
amplitude is observed, retrieving the energy consumed by the system which is
no more than the integral of the load duration distribution.
As shown in Fig.~\ref{fig09}, similarly to what is observed for the RFC,
the LDD is also properly reproduced for sufficiently sampled loads (largest
values).

The analysis presented in this work
does not take into account the wake effects, which
are very important for fatigue loads in wind turbines placed in large
off-shore farms. As one sees in Fig.~\ref{fig02}, from the main wind
directions during January 2013 turbine AV5 lies within the wake of
AV4.
Still, our method was able to accurately predict the fatigue load
observed in AV5. 
This can be explained by noticing that
the evolution of the torque in AV5 is conditioned to the wind velocity
measurements taken at its nacelle anemometer, which is able to capture
the observed turbulence intensity, even under the wake effect from
turbine AV4.

\section{Conclusions}
\label{sec:discuss}

In this paper we applied a framework to one single turbine
of Alpha Ventus wind farm for reproducing the loads observed in 
other similar wind turbines of the same wind farm. The framework
consists of the derivation of a stochastic evolution equation
for the loads constrained to the wind speed observed at each 
wind turbine.
Using rainflow counting methods and load duration distributions 
to empirical
and modeled data we show that our framework is able to accurately 
estimate fatigue loads.
In a more general context, this procedure can be applied to other
properties one wants to access for monitoring and controlling
wind turbines.

We stress that the coefficients used to reconstruct the torque at
the second turbine are obtained only from the torque and wind
velocity at the first turbine. The torque reconstruction is then performed
by taken the wind velocity at the second turbine as input parameter.
Notice that the
coefficients are taken directly from the numerical derivation
(Figs.~\ref{fig05}) in only one realization. This is possible due to the coupling
between $v$ and $T$ which is assumed to be the same for both turbines.

Two points should be stressed for future work. 
First, the conditional Langevin model here presented and applied has
some limitations. Since it is based on the integration of a stochastic
differential equation of the load, using empirical input -- the wind 
velocity -- in time periods during which no measurements are available,
additional assumptions would have to be
taken for reconstructing the torque during 
that period. 
Moreover, our analysis assumes that the wind velocity field
to which the turbine reacts can be completely described by the single
anemometer measurements; though such assumption can be taken as a first
approximation to the wind velocity field, some properly defined 
rotor-effective
velocity may improve the reconstruction in our model.

Second, having shown the ability of our method to predict one
particular load type, namely the one on the drive train, a systematic
study of how it works for the rest of the loads on a wind turbine
shall proceed.
These points will be addressed in a future work.


\acknowledgments{Acknowledgments}

The authors thank Philip Rinn and David Bastine for useful
discussions.
This work is funded by the German Environment Ministry as part of the
research project ``Probabilistic loads description, monitoring, and
reduction for the next generation offshore wind turbines (OWEA Loads)''
under grant number 0325577B.
The authors also thank Senvion SE for providing the data here analyzed.

\authorcontributions{Author Contributions}

PGL and IH performed the simulations.
PGL and MW prepared the manuscript.
JP proposed the ideas and methodologies here proposed.
All authors revised the text and output results.

\conflictofinterests{Conflicts of Interest}

The authors declare no conflict of interest. 

\bibliography{LindWaechterPeinke}
\bibliographystyle{mdpi}

\end{document}